\documentclass[preprint,
superscriptaddress,
showpacs,
 amsmath,amssymb,
 aps,
pra,
]{revtex4-1}

\usepackage{graphicx}
\usepackage{dcolumn}
\usepackage{bm}
\usepackage{color}
\usepackage[bottom]{footmisc}
\usepackage{ulem}
\usepackage{setspace}

\usepackage[nottoc]{tocbibind}
\usepackage{blindtext}
\usepackage{titlesec}

\begin{document}


\title{Inter-transition interference in spectrum of Kerr parametric oscillators}

\author{Shumpei Masuda}
\affiliation{Global Research and Development Center for Business by Quantum-AI technology (G-QuAT), National Institute of Advanced Industrial Science and Technology (AIST), Tsukuba, Ibaraki 305-8568, Japan}
\email{shumpei.masuda@aist.go.jp}


\date{\today}

\begin{abstract}
We theoretically investigate reflection and transmission measurements of two-photon and four-photon Kerr parametric oscillators (KPOs), introducing interference effects between inter-level transitions.
Due to the level degeneracy of a KPO, a probe field can be resonant with multiple inter-level transitions.
We extend the previous theory of reflection measurements by incorporating the interaction between inter-level transitions
and off-diagonal elements of the density matrix which had previously been neglected. 
We demonstrate that interference among these transitions substantially modifies the spectrum.
We identify the conditions for the interference, as well as those under which the off-diagonal elements of the density matrix affect the spectrum.
The theory is also generalized to transmission measurements, and is applicable to a broad class of systems beyond KPOs. 
\end{abstract}

\maketitle


\section{Introduction}
A Kerr nonlinear resonator with a parametric drive, called Kerr parametric oscillator (KPO), can function as a Kerr-cat qubit when operating in the quantum regime, where the nonlinearity is greater than the photon loss rate~\cite{Cochrane1999,GotoPRA2016,Puri2017b,Goto2019}.
KPOs can be realized using Josephson parametric oscillators~\cite{Meaney2014,Wang2019,Yamaji2022,Kwon2022,Yamaji2023}, charge-driven transmons with superconducting nonlinear asymmetric inductive elements (SNAIL)~\cite{Grimm2020,Venkatraman2022,Frattini2024}, trapped ions~\cite{Ding2017}, or nonlinear mechanical oscillators~\cite{Marti2024}.
The Kerr-cat qubit has the bit-flip error much less than the phase-flip error, and this property called biased error enables us to utilize quantum-error-correction schemes with less overhead compared to other qubits with unbiased errors~\cite{Tuckett2019,Puri2020,Ataides2021}.
KPO offers a promising route to a quantum computer~\cite{GotoPRA2016,Puri2017b,Grimm2020,Xu2022,Kanao2022,Chono2022,Iyama2023,Hoshi2025} and find applications to quantum annealing \cite{Goto2016,Nigg2017,Puri2017,Zhao2018,Goto2019,Onodera2020,Goto2020a,Kewming2020,Kanao2021,Yamaji2023,Yamaji2025}
and Boltzmann sampling~\cite{Goto2018}, and offer a platform to study quantum phase transitions~\cite{Dykman2018,Rota2019,Kewming2022}, quantum tunneling~\cite{Wielinga1993,Miguel2025} and chaos~\cite{Milburn1991,Hovsepyan2016,Goto2021b}.

The successful implementation of these applications relies on the accurate extraction of system parameters.
System parameters of KPOs, such as the pump field amplitude, can be estimated by measuring the energy spectrum of the system.
Various spectroscopic methods have been proposed and demonstrated~\cite{Wang2019,Grimm2020,Masuda2021,Yamaji2022,Venkatraman2022,Yamaguchi2024,Frattini2024}.

In particular, reflection measurement for periodically driven system including KPOs was theoretically formulated~\cite{Masuda2021,Suzuki2023}, and reflection measurements of a KPO were experimentally demonstrated~\cite{Yamaguchi2024}. 
However, the previous theory~[\citenum{Masuda2021}] has the following limitations: (a) it neglects interference between distinct energy-level transitions, and (b) assumes that off-diagonal elements of density matrix vanish.
Importantly, several inter-level energies can be close to each other with typical system parameters of KPOs, due to the pairwise level degeneracy~\cite{Goto2019}.
Moreover, the zero-frequency off-diagonal elements are finite even when the KPO is in the stationary state due to decoherence as shown later.
Therefore, the modification of the theory is required to accurately obtain the reflection coefficient.  
The theory for transmission measurement is also desirable because it is routinely used in superconducting-circuit experiments.

\begin{figure}[h!]
\begin{center}
\includegraphics[width=10cm]{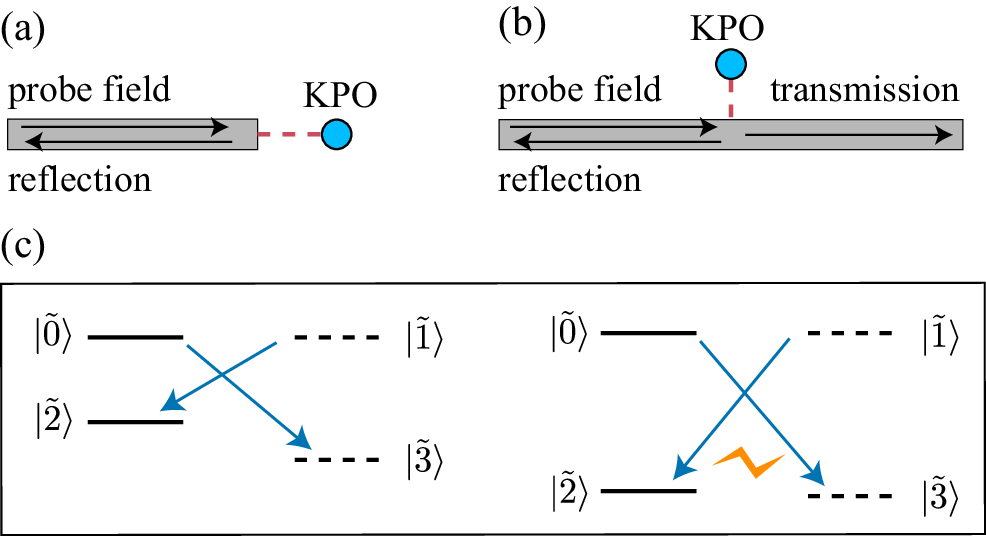}
\end{center}
\caption{
Schematic of (a) the reflection and (b) the transmission measurements. 
A KPO (blue circle) is attached to a transmission line (grey bar), where an incoming and an outgoing microwaves propagate.
The dashed lines illustrate the coupling between the KPO and the transmission line.
(c) Schematic energy diagram of a two-photon KPO showing the four relevant energy levels.  
The left panel corresponds to smaller pump amplitude, and the right panel corresponds to larger pump amplitude. 
The arrows indicate relevant transitions. 
The solid lines and the dashed lines represent energy levels with even and odd parity, respectively.
In the left panel, the transition frequencies for $|\tilde{0}\rangle \rightarrow |\tilde{3}\rangle$ and $|\tilde{1}\rangle \rightarrow |\tilde{2}\rangle$ are different, and thus the interference effect is small. In contrast, the transition frequencies are approximately the same in the right panel, resulting in significant interference (indicated by an orange zigzag line). 
}
\label{model_2_27_18}
\end{figure}

In this paper, we extend the previous theory to incorporate the interaction between inter-level transitions
and the zero-frequency off-diagonal elements of the density matrix.
Spectra of KPOs are calculated numerically and analytically.
Schematic of the reflection and transmission measurements considered in this paper are presented in figure~\ref{model_2_27_18}(a) and \ref{model_2_27_18}(b).
We show that interference effect of inter-level transitions manifests itself in the spectrum when the transitions are energetically adjacent.
Typical energy diagrams of a two-photon KPO are illustrated in figure~\ref{model_2_27_18}(c), where two inter-level transitions interfere when a probe field is applied. 
The transition energy is not only the condition for the interference between transitions. 
We clarify all the conditions of the interference.
Moreover, we show the conditions when an off-diagonal element of the density matrix affects the spectrum.
In addition, we investigate the reflection spectrum of the four-photon KPO, and develop a theory for transmission measurements of KPOs.
Our framework is broadly applicable to systems beyond KPOs. 

\section{Method}
\label{Methods}
Although our theory can be applied to broader class of periodically driven system, here we particularly consider a system composed of a two-photon KPO and a transmission line (TL) attached.
First, we consider reflection measurement illustrated in figure~\ref{model_2_27_18}(a), and later
we develop the theory for transmission measurement illustrated in figure~\ref{model_2_27_18}(b).
The KPO is driven by a pump field with the frequency of $\omega_p$.

To clarify the modifications of the theory, we first briefly review the previous one~[\citenum{Masuda2021}]. 
In a rotating frame at frequency $\omega_p/2$, the effective Hamiltonian of the system is represented as
\begin{eqnarray}
\frac{\mathcal{H}}{\hbar} &=& \frac{\mathcal{H}_{\rm KPO}^{(2)}}{\hbar} 
+   \int_{0}^{\infty} dk \Big{[} (v_b k-\omega_p/2) b_{k}^\dagger b_{k} +
 \sqrt{\frac{v_b\kappa_{\rm ex}}{2\pi}}(a^\dagger b_{k} +  b_{k}^\dagger a) \Big{]}\nonumber \\ 
&& +   \int_{0}^{\infty} dk \Big{[} (v_c k-\omega_p/2) c_{k}^\dagger c_{k} +
 \sqrt{\frac{v_c\kappa_{\rm int}}{2\pi}}(a^\dagger c_{k} +  c_{k}^\dagger a) \Big{]}.
\label{H_4_12_23}
\end{eqnarray}
The Hamiltonian of the two-photon KPO, $\mathcal{H}_{\rm KPO}^{(2)}$, is written as
\begin{eqnarray}
 \frac{\mathcal{H}_{\rm KPO}^{(2)}}{\hbar} =  \Delta a^\dagger a - \frac{K}{2}a^{\dagger 2} a^2 + \frac{p}{2}(a^2 + a^{\dagger 2}),
\label{H2KPO_6_24_25}
\end{eqnarray}
where the superscript indicate the two-photon KPO.
In equation~(\ref{H2KPO_6_24_25}), $\Delta$ is the detuning defined by $\Delta=\omega-K-\omega_p/2$; $\omega$, $p$, and $K(>0)$ are the angular resonance frequency of the KPO when no pump field is applied, the amplitude of the pump field called pump amplitude, and the nonlinearity parameter.
(We changed the notation of the pump amplitude and the nonlinearity parameter from~\cite{Masuda2021} for consistency with other papers. The pump amplitude $p$ and the nonlinearity parameter $K$ are related to the previous notations, $\beta$ and $\chi$, as $2\beta=p$ and $K=\chi$, respectively.)
In equation~(\ref{H_4_12_23}), $b_k$ and $c_k$ are annihilation operators of the eigenmodes of the TL and the loss channel with the wave number $k$, which satisfy $[b_k,b_{k'}]=\delta(k-k')$, and $[c_k,c_{k'}]=\delta(k-k')$.
Their phase velocities are denoted by $v_{b}$ and $v_c$. We assume $v_c=v_b$ in this paper.
The loss rate to the TL (loss channel) is denoted by $\kappa_{\rm ex}$ ($\kappa_{\rm int}$).
The terms with $b_k$ in equation~(\ref{H_4_12_23}) correspond to the Hamiltonian of the eigenmodes of the TL and the coupling between the modes and the KPO, while
the terms with $c_k$ correspond to the Hamiltonian of the eigenmodes representing a loss channel and their coupling to the KPO.
We note that the rotating wave approximation was applied in equation~(\ref{H_4_12_23}), namely we neglected rapidly oscillating terms with the frequency natural number multiple of $\omega_p$.
The approximation is valid when $\omega_p\gg p, \Delta, K$.

We assume that a probe field with the angular frequency of $\omega_{\rm in}$ and the amplitude of $E$ is injected to the KPO, and the reflection is measured.
The reflection coefficient is represented as~\cite{Masuda2021}
\begin{eqnarray}
\Gamma  = 1 - \frac{i}{E} \sqrt\frac{\kappa_{\rm ex}}{v_b} \sum_{mn} X_{mn} \rho^{\rm (F)}_{nm}[-\tilde\omega_{\rm in}]
\label{Gamma_6_23_25}
\end{eqnarray}
with $\tilde{\omega}_{\rm in}=\omega_{\rm in}-\omega_p/2$.
Here, $\rho^{\rm (F)}_{nm}[-\tilde\omega_{\rm in}]$ is the Fourier component of $\rho_{nm}(t)= \langle \phi_n | \rho(t) | \phi_m \rangle$ at a frequency of $-\tilde\omega_{\rm in}$,
where $\rho(t)$ is the density operator at time $t$, and $|\phi_i\rangle$ is an eigenstate of $\mathcal{H}_{\rm KPO}^{(2)}$.
$\rho^{\rm (F)}_{nm}[-\tilde\omega_{\rm in}]$ is regarded as the contribution to the reflection coefficient from the transition $|\phi_m\rangle \rightarrow |\phi_n\rangle$.
$X_{mn}$ is defined by 
\begin{eqnarray}
X_{mn} = \langle \phi_m | a | \phi_n \rangle.
\label{X_9_2_25}
\end{eqnarray}

The equation of motion of the KPO was derived as~\cite{Masuda2021}
\begin{eqnarray}
\dot\rho_{nm} &=& i(\omega_m - \omega_n) \rho_{nm} 
- i\Omega \sum_k (X_{nk} \rho_{km} - X_{km} \rho_{nk} ) e^{i\tilde\omega_{\rm in} t}\nonumber\\
&& - i\Omega \sum_k (X_{kn}^\ast \rho_{km} - X_{mk}^\ast \rho_{nk} ) e^{-i\tilde\omega_{\rm in}t}\nonumber\\
&& + \kappa_{\rm tot} \sum_{kl} X_{nl} X_{mk}^\ast \rho_{lk} 
-\frac{\kappa_{\rm tot} }{2}  \sum_{k} (Y_{nk} \rho_{km} + Y_{km} \rho_{nk} ),
\label{drho_3_18_21}
\end{eqnarray}
where 
\begin{eqnarray}
Y_{mn}=\langle \phi_m|a^\dagger a|\phi_n\rangle,
\label{Y_9_2_25}
\end{eqnarray}
and $\omega_i$ is an eigenvalue of $\mathcal{H}_{\rm KPO}^{(2)}/\hbar$; 
$\kappa_{\rm tot}=\kappa_{\rm ex}+\kappa_{\rm int}$; and
$\Omega$ is defined by 
\begin{eqnarray}
\Omega=\sqrt{v_b\kappa_{\rm ex}}E.
\end{eqnarray}
In principle, the reflection coefficient can be obtained by numerically integrating equation~(\ref{drho_3_18_21}) and calculating $\rho^{\rm (F)}_{nm}[-\tilde\omega_{\rm in}]$ determining the reflection coefficient in equation~(\ref{Gamma_6_23_25}). 
However, performing this calculation for various values of $\tilde\omega_{\rm in}$ and system parameters is computationally expensive and time-consuming.
In order to obtain $\rho^{\rm (F)}_{nm}[-\tilde\omega_{\rm in}]$,
we consider the Fourier transform of equation (\ref{drho_3_18_21}) with a frequency of $-\tilde\omega_{\rm in}$ written as
\begin{eqnarray}
0 &=& i(\tilde\omega_{\rm in} + \omega_m - \omega_n)\rho^{\rm (F)}_{nm}[-\tilde\omega_{\rm in}]
- i\Omega \sum_k (X_{nk} \rho^{\rm (F)}_{km}[-2\tilde\omega_{\rm in}] - X_{km} \rho^{\rm (F)}_{nk}[-2\tilde\omega_{\rm in}] )\nonumber\\
&& - i\Omega \sum_k (X_{kn}^\ast \rho^{\rm (F)}_{km}[0] - X_{mk}^\ast \rho^{\rm (F)}_{nk}[0] ) 
+ \kappa_{\rm tot} \sum_{kl} X_{nl} X_{mk}^\ast \rho^{\rm (F)}_{lk} [-\tilde\omega_{\rm in}]
\nonumber\\
&&  -\frac{\kappa_{\rm tot} }{2}  \sum_{k} (Y_{nk} \rho^{\rm (F)}_{km}[-\tilde\omega_{\rm in}] + Y_{km} \rho^{\rm (F)}_{nk}[-\tilde\omega_{\rm in}] ).
\label{rho_F_v1_2_20_21}
\end{eqnarray}

\subsection{Previous method}
In the prior studies~\cite{Masuda2021,Suzuki2023}, it was assumed that the off-diagonal elements of the density matrix at zero frequency, $\rho^{\rm (F)}_{mn(\ne m)}[0]$, are negligible, and that $\rho^{\rm (F)}_{lk}[-\tilde\omega_{\rm in}]=0$ unless $k=m$ and $l=n$.
The later assumption corresponding to the case that the input field is resonant only with the transition from $|\phi_m\rangle$ to $|\phi_n\rangle$. 
Then, equation~(\ref{rho_F_v1_2_20_21}) leads to the formula for $\rho^{\rm (F)}_{nm}[-\tilde\omega_{\rm in}]$, presented as
\begin{eqnarray}
{\rho^{\rm (F)}}_{nm}[-\tilde\omega_{\rm in}] = \frac{i\Omega X_{mn}^\ast (\rho^{\rm (F)}_{mm}[0] - \rho^{\rm (F)}_{nn}[0]) }{i\Delta_{nm}- \frac{\kappa_{\rm ex}+\kappa_{\rm int}}{2}(Y_{mm}+Y_{nn})},
\label{rhomn_2_19_21}
\end{eqnarray}
with $\Delta_{nm} = \tilde\omega_{\rm in}  - \omega_n + \omega_m$.
Equations~(\ref{Gamma_6_23_25}) and (\ref{rhomn_2_19_21}) were used to calculate the reflection coefficient previously.
However, the first assumption holds only for special states of the KPO; 
the second approximation can be invalid when relevant transition frequencies are close to each other, as multiple matrix elements, $\rho^{\rm (F)}_{lk}[-\tilde\omega_{\rm in}]$, may interfere. 
The interference can have a significant impact on the reflection coefficient, as will be presented later.

\subsection{Modified method}
\label{Modified method}
In this paper, we eliminate the previous assumptions and highlight the significance of the factors previously disregarded. 
Meanwhile, we still consider the weak-input-field limit; that is, we assume that the effect of the probe field on the density matrix is treated perturbatively up to first order in $\Omega$.
We consider the case where the KPO is in a stationary state in the absence of a probe field. In this case, with respect to the order in $\Omega$, $\rho^{\rm (F)}_{lk}[0]$ is $\mathcal{O}(1)$, while $\rho^{\rm (F)}_{lk}[l\tilde\omega_{\rm in}]$ is $\mathcal{O}(\Omega^{n(\ge 1)})$ for $|l|>0$.

Equation~(\ref{rho_F_v1_2_20_21}) can be rewritten as 
\begin{eqnarray}
0 &=& i\Delta_{nm} \frac{\rho^{\rm (F)}_{nm}[-\tilde\omega_{\rm in}]}{\Omega}
 - i \sum_k (X_{kn}^\ast \rho^{\rm (F)}_{km}[0] - X_{mk}^\ast \rho^{\rm (F)}_{nk}[0] ) 
+ \kappa_{\rm tot} \sum_{kl} X_{nl} X_{mk}^\ast \frac{\rho^{\rm (F)}_{lk} [-\tilde\omega_{\rm in}]}{\Omega}
\nonumber\\
&&  -\frac{\kappa_{\rm tot} }{2}  \sum_{k} \Big{(} Y_{nk} \frac{\rho^{\rm (F)}_{km}[-\tilde\omega_{\rm in}]}{\Omega} + Y_{km} \frac{\rho^{\rm (F)}_{nk}[-\tilde\omega_{\rm in}]}{\Omega} \Big{)},
\label{rho_nm_2_16_22}
\end{eqnarray}
where the terms involving $\Omega \rho^{\rm (F)}_{lk}[-2\tilde\omega_{\rm in}]$ were omitted because they are of higher order in $\Omega$. 
Equation~(\ref{rho_nm_2_16_22}) implies that $\rho^{\rm (F)}_{lk}[-\tilde\omega_{\rm in}]$ is $\mathcal{O}(\Omega^1)$, because $\rho^{\rm (F)}_{lk}[0]$ is $\mathcal{O}(1)$.
Matrix element $\rho^{\rm (F)}_{nm}[-\tilde\omega_{\rm in}]$ depends on the other elements $\rho^{\rm (F)}_{lk}[-\tilde\omega_{\rm in}]$. 
This indicates the presence of interference between different transitions, noting that $\rho^{\rm (F)}_{nm}[-\tilde\omega_{\rm in}]$ can be regarded as the contribution to the reflection coefficient from the transition $|\phi_m\rangle \rightarrow |\phi_n\rangle$, as mentioned earlier.
In contrast, in the previous method, $\rho^{\rm (F)}_{nm}[-\tilde\omega_{\rm in}]$ is independent of $\rho^{\rm (F)}_{lk}[-\tilde\omega_{\rm in}]$ as seen in equation~(\ref{rhomn_2_19_21}), and thus the interference was neglected.

We obtain $\rho^{\rm (F)}_{nm}[-\tilde\omega_{\rm in}]/\Omega$ by numerically or analytically solving the simultaneous equation~(\ref{rho_nm_2_16_22}).
We use the same value of $\rho^{\rm (F)}_{kl}[0]$ as when the probe field is absent.
In the numerical calculation, we typically consider between ten and twenty relevant energy levels and neglect other excited states.
The reflection coefficient is obtained by substituting the obtained value of $\rho^{\rm (F)}_{nm}[-\tilde\omega_{\rm in}]/\Omega$ to 
$\Gamma  = 1 - i\kappa_{\rm ex} \sum_{mn} X_{mn}\rho^{\rm (F)}_{nm}[-\tilde\omega_{\rm in}]/\Omega$, 
which is equivalent to equation~(\ref{Gamma_6_23_25}).

\subsection{Analytic results}
\label{Analytic results}
We derive an analytic formula for the reflection coefficient in the case where two transitions interfere.
A KPO exhibits pairwise level degeneracy when the pump amplitude is sufficiently large. 
We consider the case where two ground states, $|\tilde{0}\rangle$ and $|\tilde{1}\rangle$, and a pair of excited states, $|\tilde{2}\rangle$ and $|\tilde{3}\rangle$, are each degenerate, as illustrated in figure~\ref{model_2_27_18}(c).
This pairwise level degeneracy was experimentally observed in two-photon KPOs~\cite{Frattini2024}.
Here, $|\tilde{n}\rangle$ denotes the eigenstate of $\mathcal{H}_{\rm KPO}^{(2)}$ in equation~(\ref{H_4_12_23}) that continuously evolves from Fock state $|n\rangle$ as $p$ is adiabatically increased from zero, whereas $|\phi_m\rangle$ simply denotes  an eigenstate of $\mathcal{H}_{\rm KPO}^{(2)}$.
The probe field is nearly resonant with the transitions between the ground states and the excited states, namely, $\tilde{\omega}_{\rm in}\simeq \omega_{\tilde{3}(\tilde{2})}-\omega_{\tilde{0}(\tilde{1})}$. 
Under this condition, two transitions, $|\tilde{0}\rangle\rightarrow |\tilde{3}\rangle$ and $|\tilde{1}\rangle\rightarrow |\tilde{2}\rangle$, interfere. 
(Transitions between states with the same parity, $|\tilde{0}\rangle\rightarrow |\tilde{2}\rangle$ and $|\tilde{1}\rangle\rightarrow |\tilde{3}\rangle$, can be neglected.)

Assuming that the effects of the other energy levels are negligible, we rewrite equation~(\ref{rho_nm_2_16_22}) as
\begin{eqnarray}
\left(
\begin{array}{cl}
a_{11} & a_{12} \\
a_{21} & a_{22}
\end{array}
\right)
\left(\begin{array}{c}
\rho^{\rm (F)}_{\tilde{3}\tilde{0}}[-\tilde\omega_{\rm in}]/\Omega\\
\rho^{\rm (F)}_{\tilde{2}\tilde{1}}[-\tilde\omega_{\rm in}]/\Omega
\end{array}
\right)
=
\left(\begin{array}{c}
b_1\\
b_2
\end{array}\right),
\label{eq_a_b_7_31_25}
\end{eqnarray}
where $a_{11} = i\Delta_{\tilde{3}\tilde{0}}-\frac{\kappa_{\rm tot}}{2}(Y_{\tilde{0}\tilde{0}}+Y_{\tilde{3}\tilde{3}})$,
$a_{12} = \kappa_{\rm tot}X_{\tilde{3}\tilde{2}}X^\ast_{\tilde{0}\tilde{1}}$
$a_{21} = \kappa_{\rm tot}X_{\tilde{2}\tilde{3}}X^\ast_{\tilde{1}\tilde{0}}$,
$a_{22} = i\Delta_{\tilde{2}\tilde{1}}-\frac{\kappa_{\rm tot}}{2}(Y_{\tilde{1}\tilde{1}}+Y_{\tilde{2}\tilde{2}})$,
$b_1 = i \sum_k \big{(}X_{\tilde{k}\tilde{3}}^\ast \rho^{\rm (F)}_{\tilde{k}\tilde{0}}[0] - X_{\tilde{0}\tilde{k}}^\ast \rho^{\rm (F)}_{\tilde{3}\tilde{k}}[0] \big{)}$, and
$b_2 = i \sum_k \big{(}X_{\tilde{k}\tilde{2}}^\ast \rho^{\rm (F)}_{\tilde{k}\tilde{1}}[0] - X_{\tilde{1}\tilde{k}}^\ast \rho^{\rm (F)}_{\tilde{2}\tilde{k}}[0] \big{)}$.
From equation~(\ref{eq_a_b_7_31_25}), we derive
\begin{eqnarray}
\left(\begin{array}{c}
\rho^{\rm (F)}_{\tilde{3}\tilde{0}}[-\tilde\omega_{\rm in}]/\Omega\\
\rho^{\rm (F)}_{\tilde{2}\tilde{1}}[-\tilde\omega_{\rm in}]/\Omega
\end{array}
\right)=
\frac{1}{A} \left(\begin{array}{c}
a_{22}b_1-a_{12}b_2\\
-a_{21}b_1+a_{11}b_2
\end{array}\right),
\label{rho_two_7_31_25}
\end{eqnarray}
where $A=a_{11}a_{22}-a_{12}a_{21}$.
Substituting equation~(\ref{rho_two_7_31_25}) into equation~(\ref{Gamma_6_23_25}), we obtain the reflection coefficient as
\begin{eqnarray}
\Gamma = 1-i\frac{\kappa_{\rm ex}}{A} \Big{[}
X_{\tilde{0}\tilde{3}} (a_{22}b_1-a_{12}b_2)
+ X_{\tilde{1}\tilde{2}}( a_{11}b_2-a_{21}b_1)  \Big{]}.
\label{Gamma_0312_7_24_25}
\end{eqnarray}

We now focus on the specific case where the pump amplitude is sufficiently large and the detuning, $\Delta$, is zero.
Then, the four relevant energy levels can be expressed approximatly as 
$|\tilde{0}\rangle=(D(\alpha)|0\rangle + D(-\alpha)|0\rangle)/\sqrt{2}$, $|\tilde{1}\rangle=(D(\alpha)|0\rangle - D(-\alpha)|0\rangle)/\sqrt{2}$, $|\tilde{2}\rangle=(D(\alpha)|1\rangle - D(-\alpha)|1\rangle)/\sqrt{2}$, and 
$|\tilde{3}\rangle=(D(\alpha)|1\rangle + D(-\alpha)|1\rangle)/\sqrt{2}$, with the displacement operator defined by 
$D(\alpha)=\exp(\alpha a^\dagger - \alpha^\ast a)$.
Here, $\alpha=\sqrt{p/K}$.
When the pump amplitude is sufficiently large, $|\tilde{0} (\tilde{2})\rangle$ and $|\tilde{1} (\tilde{3})\rangle$ become degenerate, and thus we have $\Delta_{\tilde{3}\tilde{0}}=\Delta_{\tilde{2}\tilde{1}}=:\tilde{\Delta}$. 
We have $\langle\alpha|-\alpha\rangle=0$ for sufficiently large $\alpha$. 
We also have $\rho^{\rm (F)}_{\tilde{0}\tilde{0}}[0]=\rho^{\rm (F)}_{\tilde{1}\tilde{1}}[0]=1/2$, while the other elements $\rho^{\rm (F)}_{\tilde{i}\tilde{j}}[0]$ are negligible for the stationary state in this parameter regime~\cite{Masuda2021}.
Using these equations in equation~(\ref{Gamma_0312_7_24_25}), we can obtain the simple expression of the reflection coefficient as
\begin{eqnarray}
\Gamma = 1 + \frac{\kappa_{\rm ex}}{i\tilde{\Delta}-\frac{\kappa_{\rm ex}+\kappa_{\rm int}}{2}}.
\label{Gamma_ana_7_24_25}
\end{eqnarray}
Interestingly, there is no $\alpha$ dependence in $\Gamma$ in the large-pump-amplitude limit, and  $\Gamma$ has the same form as that for a linear resonator, represented as
$\Gamma_r=1+\kappa_{\rm ex}^{(r)}/[i\Delta_r - (\kappa_{\rm ex}^{(r)} + \kappa_{\rm int}^{(r)} )/2 ]$,
where $\kappa_{\rm ex}^{(r)}$ and $\kappa_{\rm int}^{(r)}$ are the external and the internal loss rates;.$\Delta_r=\omega_{\rm in}-\omega_0$; $\omega_0$ is the angular resonance frequency of the resonator. 
Therefore, the nominal external loss rate and nominal internal loss rate~\cite{Masuda2021,Yamaguchi2024} are the same as those of the linear resonator.

If we neglect the interference effect, namely by setting $a_{12}=a_{21}=0$, the reflection coefficient is represented as $\Gamma' = 1 + {\kappa_{\rm ex}}/[i\tilde{\Delta}-(\kappa_{\rm ex}+\kappa_{\rm int})(2\alpha^2+1)/2]$, thus it has $\alpha$ dependence.
From the expressions for $\Gamma$ and $\Gamma'$, we observe that $|\Gamma(\tilde{\Delta}=0)|<|\Gamma'(\tilde{\Delta}=0)|$, indicating that the dip in $|\Gamma|$ becomes deeper due to the interference effect.

\subsection{Conditions for interference}
\label{Conditions for interference}
As discussed in section~\ref{Modified method}, $\rho^{\rm (F)}_{nm}[-\tilde\omega_{\rm in}]$ can depend on other element $\rho^{\rm (F)}_{ts}[-\tilde\omega_{\rm in}]$, and this dependence is interpreted as interference between transitions $|\phi_m\rangle \rightarrow |\phi_n\rangle$ and $|\phi_s\rangle \rightarrow |\phi_t\rangle$.
We discuss the conditions for the interference using equation~(\ref{rho_nm_2_16_22}).

The interference arises from the three terms, $X_{nt} X_{ms}^\ast \rho^{\rm (F)}_{ts} [-\tilde\omega_{\rm in}]$, $Y_{nt} \rho^{\rm (F)}_{tm}[-\tilde\omega_{\rm in}]$, and $Y_{sm} \rho^{\rm (F)}_{ns}[-\tilde\omega_{\rm in}]$.
Each term respectively leads to a condition for the interaction between $\rho^{\rm (F)}_{nm}[-\tilde\omega_{\rm in}]$ and $\rho^{\rm (F)}_{ts}[-\tilde\omega_{\rm in}]$:
(i) $\mathcal{P}(\phi_t)\ne \mathcal{P}(\phi_n)$ and $\mathcal{P}(\phi_s)\ne \mathcal{P}(\phi_m)$;
(ii) $s=m$ and $\mathcal{P}(\phi_t)=\mathcal{P}(\phi_n)$;
(iii) $t=n$ and $\mathcal{P}(\phi_s)=\mathcal{P}(\phi_m)$,
where $\mathcal{P}(\phi_i)$ denotes the parity of $|\phi_i\rangle$.
The interference occurs when at least one of these conditions is satisfied.
The conditions arise from the definitions of $X_{ji}$ and $Y_{ji}$ given in equations~(\ref{X_9_2_25}) and~(\ref{Y_9_2_25}), as well as from the fact that parity is conserved under the Hamiltonian of the KPOs considered in this study. Throughout this paper, states with even indices are assigned even parity, whereas states with odd indices are assigned odd parity.


The following conditions should also be satisfied.
(iv) The probe frequency $\tilde{\omega}_{\rm in}$ is close to both transition frequencies $\omega_n-\omega_m$ and $\omega_t-\omega_s$;
(v) relevant density matrix elements at zero frequency, such as $\rho^{\rm (F)}_{km}[0]$ and $\rho^{\rm (F)}_{nk}[0]$, are sufficiently large.
Unless condition (iv) is satisfied, large value of $\Delta_{nm} (\Delta_{ts})$ suppresses
$\rho^{\rm (F)}_{nm} [-\tilde\omega_{\rm in}] (\rho^{\rm (F)}_{ts} [-\tilde\omega_{\rm in}])$.
Condition (v) is also necessary so that both $\rho^{\rm (F)}_{nm} [-\tilde\omega_{\rm in}]$ and $\rho^{\rm (F)}_{ts} [-\tilde\omega_{\rm in}]$ are sufficiently large. 

\subsection{Effect of zero-frequency off-diagonal element $\rho^{\rm (F)}_{ts(\ne t)}[0]$}
We discuss conditions under which zero-frequency off-diagonal element $\rho^{\rm (F)}_{ts(\ne t)}[0]$ influences $\rho^{\rm (F)}_{nm}[-\tilde\omega_{\rm in}]$, using equation~(\ref{rho_nm_2_16_22}). 
Terms, $X_{tn}^\ast\rho_{tm}^{\rm (F)}[0]$ and $-X_{ms}^\ast\rho_{ns}^{\rm (F)}[0]$, imply that off-diagonal element $\rho^{\rm (F)}_{ts(\ne t)}[0]$ can affect $\rho^{\rm (F)}_{nm}[-\tilde\omega_{\rm in}]$ 
if $s=m$ and $\mathcal{P}(\phi_t)\ne \mathcal{P}(\phi_n)$, or
if $t=n$ and $\mathcal{P}(\phi_s)\ne \mathcal{P}(\phi_m)$.
The conditions on the parity of states come from the factor $X_{tn}^\ast$ and $X_{ms}^\ast$.
As an example, this analysis suggests that $\rho^{\rm (F)}_{30}[-\tilde\omega_{\rm in}]$, corresponding to transition $|\phi_0\rangle\rightarrow |\phi_3\rangle$, depends on $\rho^{\rm (F)}_{20}[0]$.

\section{Numerical results}
\label{Numerical results}
\begin{figure}[h!]
\begin{center}
\includegraphics[width=13cm]{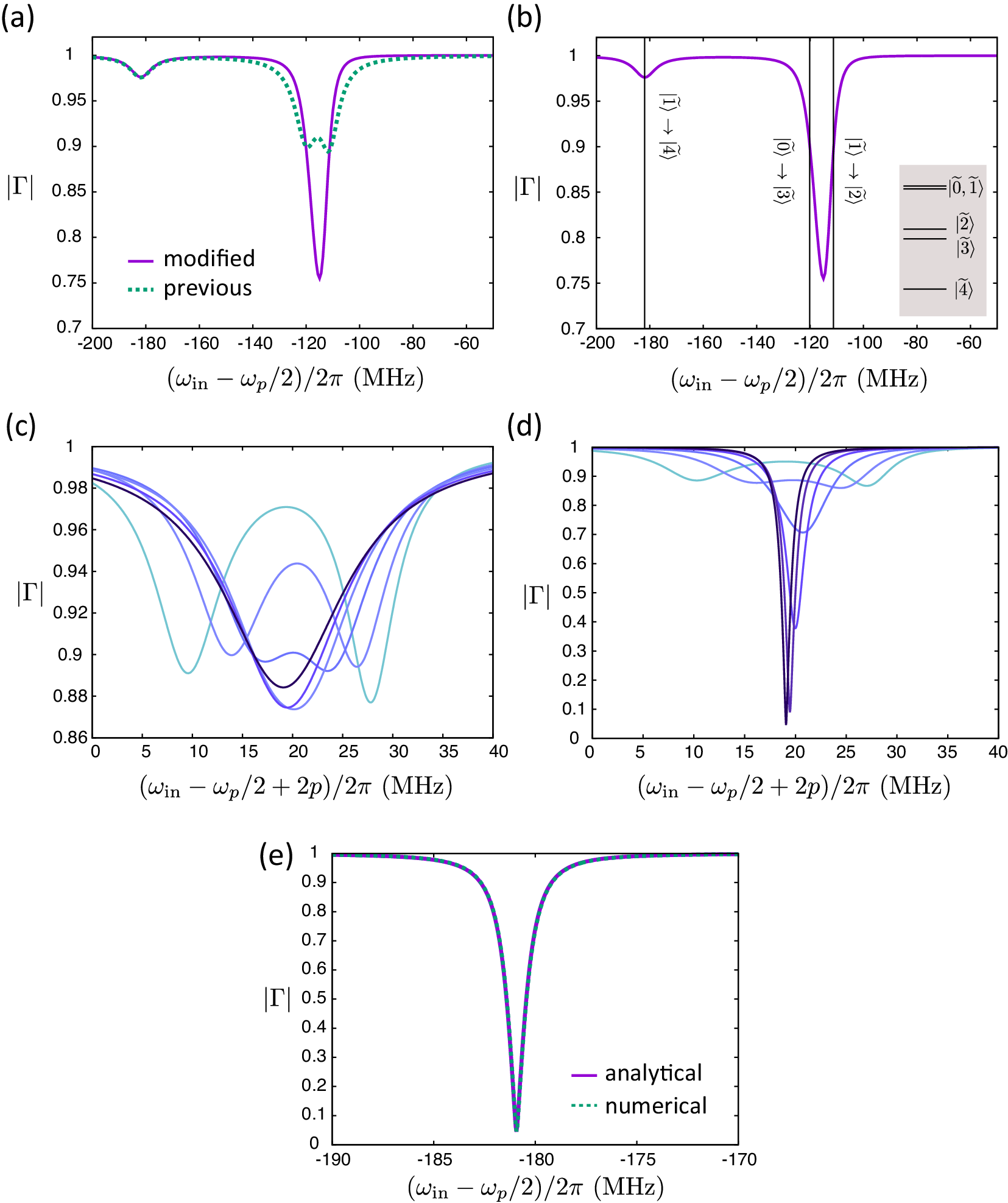}
\end{center}
\caption{
(a) Amplitude of the reflection coefficient as a function of $\omega_{\rm in}$ for a two-photon KPO in the stationary state. 
The purple solid (green dashed) curve corresponds to the modified (previous) theory.
We used $p/2\pi=68$~MHz, $\Delta/2\pi=0$~MHz, $K/2\pi=17$~MHz, $\kappa_{\rm ex}/2\pi=1$~MHz, and $\kappa_{\rm int}/2\pi=0.45~$MHz.
Panel (b) shows the transition frequencies between energy levels indicated by the vertical lines overlaid on the data presented in panel (a).
The inset illustrates five relevant energy levels. 
Amplitude of the reflection coefficient as a function of $\omega_{\rm in}$ for the previous theory (c) and for the modified theory (d).
The pump amplitude, $p$, ranges from $50~{\rm MHz}$ to $100~{\rm MHz}$ in steps of 10~{\rm MHz}.
The darker color of the curves denote larger pump amplitude.
The other parameters are the same as panel (a).
The factor $2p$ in the horizontal axis is introduced to cancel the shift of dips as $p$ is changed.
(e) Amplitude of the reflection coefficient as a function of $\omega_{\rm in}$ for $p/2\pi=100$~MHz. The analytical result in equation~(\ref{Gamma_0312_7_24_25}) overlaps with the numerical result.
}
\label{spectrum_slice_3_10_25}
\end{figure}

The theory is applied to the reflection measurement of a two-photon KPO in order to analyze the effect of the interference.
We assume that the KPO is in the stationary state.
More precisely, the value of $\rho^{\rm (F)}_{kl}[0]$ in equation~(\ref{rho_nm_2_16_22}) is taken to be the same as that of the stationary state in the absence of the probe field. 
Figure~\ref{spectrum_slice_3_10_25}(a) compares the amplitude of the reflection coefficient calculated using the modified theory and the previous one.
Figure~\ref{spectrum_slice_3_10_25}(b) shows the transition frequencies between energy levels.
Relevant energy levels, $|\tilde{n}\rangle$, are illustrated in the inset.
The parameter set used for figure~\ref{spectrum_slice_3_10_25}(a,b) corresponds to the large-pump-amplitude regime discussed in section~\ref{Analytic results}, where the transition frequencies of $|\widetilde{0}\rangle\rightarrow |\widetilde{3}\rangle$ and $|\widetilde{1}\rangle\rightarrow |\widetilde{2}\rangle$ are close to each other. 
This pair of transitions satisfies the condition~(i) for the interference.
As predicted in section~\ref{Analytic results}, 
the modified theory exhibits a sharper and deeper dip in $|\Gamma|$ near the transition frequencies compared to the previous one, due to the interference effect.
To highlight the interference effect, we calculate $|\Gamma|$ for various values of the pump amplitude $p$ while keeping the other parameters fixed to those used in figure~\ref{spectrum_slice_3_10_25}(a,b).
As the pump amplitude increases, $|\tilde{2}\rangle$ and $|\tilde{3}\rangle$ become degenerate, as illustrated in figure~\ref{model_2_27_18}(c).
The two distinct dips in $|\Gamma|$, corresponding to $|\tilde{0}\rangle\rightarrow |\tilde{3}\rangle$ and $|\tilde{1}\rangle\rightarrow |\tilde{2}\rangle$, begin to merge as their associated frequencies approach each other, as shown in figure~\ref{spectrum_slice_3_10_25}(c,d).  
In the modified theory, the merged dip is sharp and deep, in contrast to the previous one. 
Figure~\ref{spectrum_slice_3_10_25}(e) compares the analytical result in equation~(\ref{Gamma_0312_7_24_25}) with the numerical result.
The analytical result agrees well with the numerical one.

\begin{figure}[h!]
\begin{center}
\includegraphics[width=13cm]{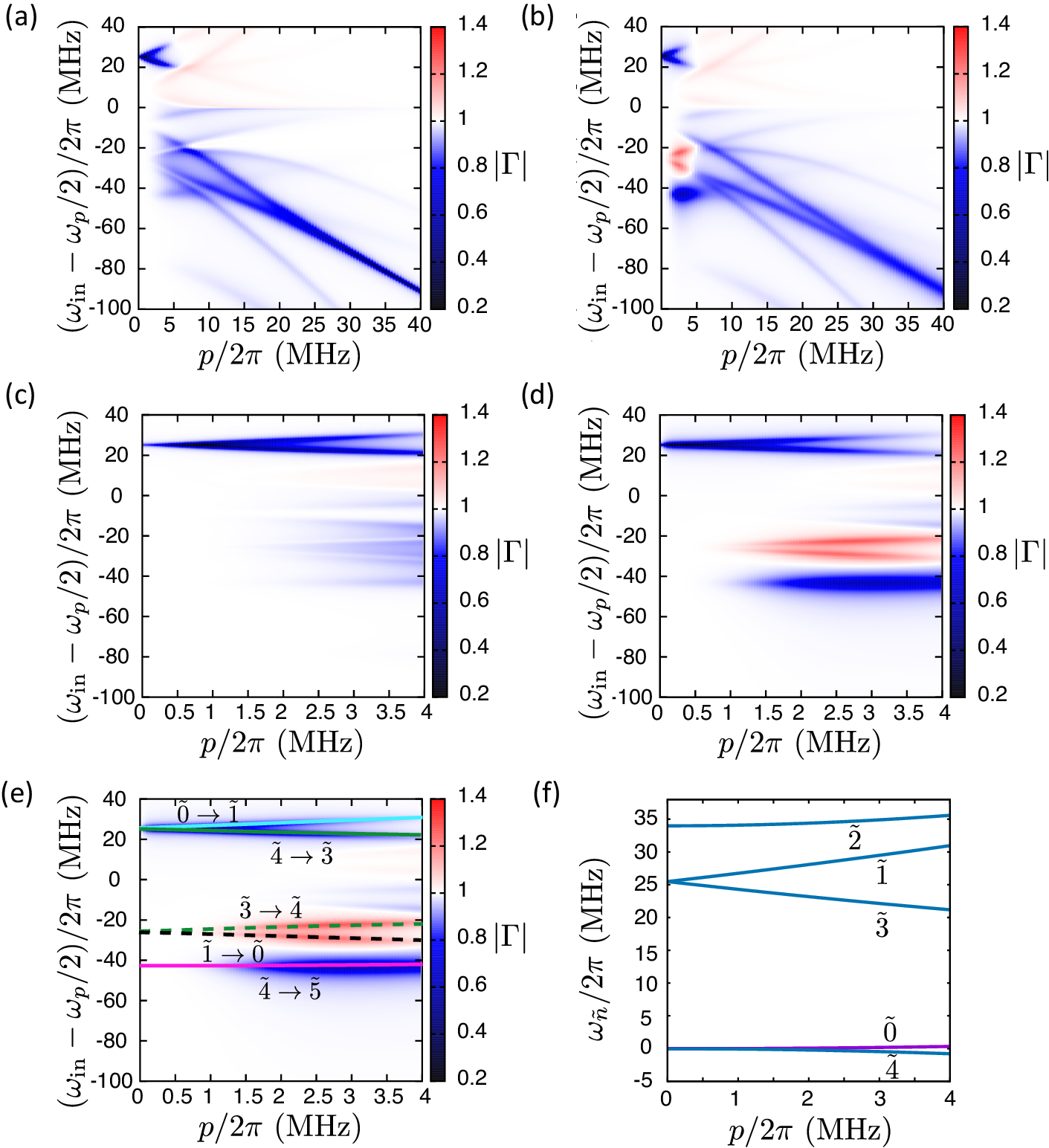}
\end{center}
\caption{
Amplitude of the reflection coefficient as a function of $\omega_{\rm in}$ and $p$ for a two-photon KPO in the stationary state.
The parameters used are $K/2\pi=17$~MHz,  $\Delta=1.5K$, $\kappa_{\rm ex}/2\pi=1$~MHz, and $\kappa_{\rm int}/2\pi=0.1~$MHz.
Panels (a) and (b) correspond to the modified theory and the previous one, respectively.
Panels (c) and (d) show the same data as panels (a) and (b), but over a shorter range of $p$. 
Panel (e) shows the frequencies of relevant transitions overlaid on the data presented in panel (d). 
Panel (f) presents the energy diagram of the system.
Letter $\tilde{n}$ on the figure indicates the energy eigenstate $|\tilde{n}\rangle$.
}
\label{fourier_com_6_8_25}
\end{figure}

\begin{figure}[h!]
\begin{center}
\includegraphics[width=7.5cm]{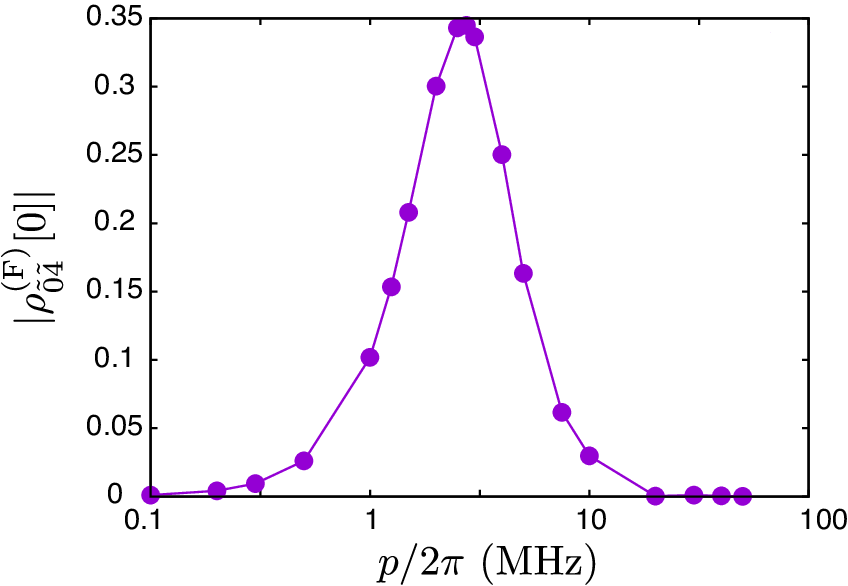}
\end{center}
\caption{
Amplitude of the density matrix element $|\rho^{\rm (F)}_{\tilde{0}\tilde{4}}[0]|(=|\rho^{\rm (F)}_{\tilde{4}\tilde{0}}[0]|)$ of the stationary state as a function of $p$.
The parameters are the same as in figure~\ref{fourier_com_6_8_25}.
The solid line is shown as a guide to the eye.
}
\label{rho_p_dep_2p_6_11_25}
\end{figure}

Next, we demonstrate that the reflection spectrum depends on the zero-frequency off-diagonal elements of the density matrix analyzing the regime of small pump amplitude, $p<K$.
Figure~\ref{fourier_com_6_8_25}(a-d) show the amplitude of the reflection coefficient as a function of $\omega_{\rm in}$ and $p$.
The parameters used are given in the caption of the figure. 
Figure~\ref{fourier_com_6_8_25}(a) is for the modified theory, and figure~\ref{fourier_com_6_8_25}(b) is for the previous one, in which the effect of off-diagonal elements of the density matrix is neglected.
Figures~\ref{fourier_com_6_8_25}(c) and \ref{fourier_com_6_8_25}(d) show the same data as figures~\ref{fourier_com_6_8_25}(a) and \ref{fourier_com_6_8_25}(b), but over a shorter range of $p$.
They show a difference, especially in the spectrum corresponding to the transitions, $|\tilde{3}\rangle\rightarrow |\tilde{4}\rangle,~|\tilde{1}\rangle \rightarrow |\tilde{0}\rangle$, and $|\tilde{4}\rangle \rightarrow |\tilde{5}\rangle$, indicated in figure~\ref{fourier_com_6_8_25}(e).
With the parameters used, $|\tilde{1}\rangle$ and $|\tilde{3}\rangle$, as well as $|\tilde{0}\rangle$ and $|\tilde{4}\rangle$, are degenerate at $p=0$, as shown in figure~\ref{fourier_com_6_8_25}(f).
We attribute the difference between figure~\ref{fourier_com_6_8_25}(a,c) and figure~\ref{fourier_com_6_8_25}(b,d) to the effect of the zero-frequency off-diagonal elements of the density matrix, because the modified theory yields a result similar to that of the previous theory when we set $\rho^{\rm (F)}_{\tilde{k}\tilde{l}(\ne \tilde{k})}[0]=0$.
Particularly, $|\rho^{\rm (F)}_{\tilde{0}\tilde{4}}[0]|$ is considerably large within the range of $p$ used in figure~\ref{fourier_com_6_8_25}(c,d), as shown in figure~\ref{rho_p_dep_2p_6_11_25}.

\section{Four-photon Kerr parametric oscillator}
Four-photon KPOs have been attracting attention for its potential to implement autonomous quantum error correction~\cite{Kwon2022}.
In this section, we analyze the reflection measurement of a four-photon KPO.

\begin{figure}[h!]
\begin{center}
\includegraphics[width=7.5cm]{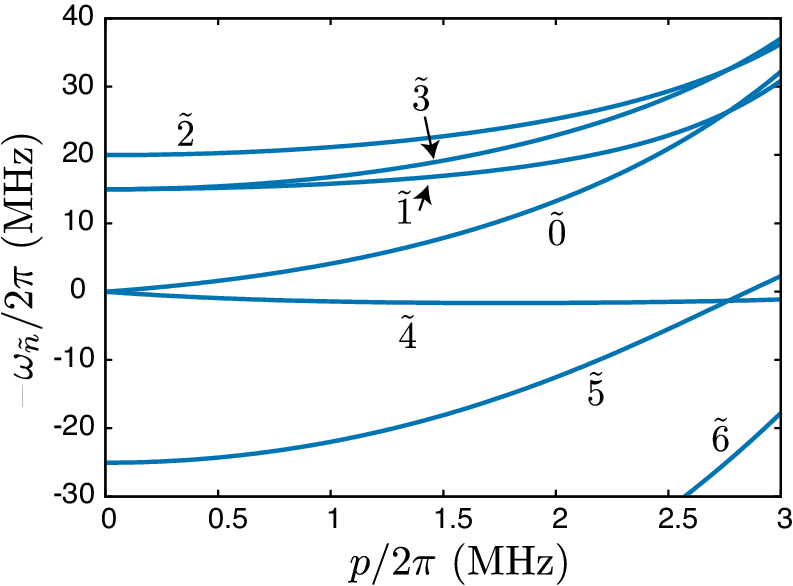}
\end{center}
\caption{
Energy diagram of a four-photon KPO as a function of $p$.
The parameter set used is $K/2\pi=10$~MHz and $\Delta=1.5K$.
Letter $\tilde{n}$ on the figure indicates the energy eigenstate $|\tilde{n}\rangle$.
}
\label{eig_dif_5_12_25}
\end{figure}

The effective Hamiltonian of a four-photon KPO in a rotating frame at frequency $\omega_p/4$ is represented as~\cite{Kwon2022}
\begin{eqnarray}
\frac{\mathcal{H}_{\rm KPO}^{\rm (4)}}{\hbar} =&=& \Delta a^\dagger a - \frac{K}{2}a^{\dagger 2} a^2 + \frac{p}{2}(a^{\dagger 4}+a^4),
\end{eqnarray}
where $\omega_p$ is the angular frequency of a four-photon pump field.
The energy diagram of the KPO is presented in figure~\ref{eig_dif_5_12_25}.
Here, $|\tilde{n}\rangle$ denotes the eigenstate of $\mathcal{H}_{\rm KPO}^{(4)}$ that continuously evolves from Fock state $|n\rangle$ as $p$ is adiabatically increased from zero. 
We set the detuning to $\Delta=1.5K$, as in Ref.~\cite{Kwon2022}.

\begin{figure}[h!]
\begin{center}
\includegraphics[width=13cm]{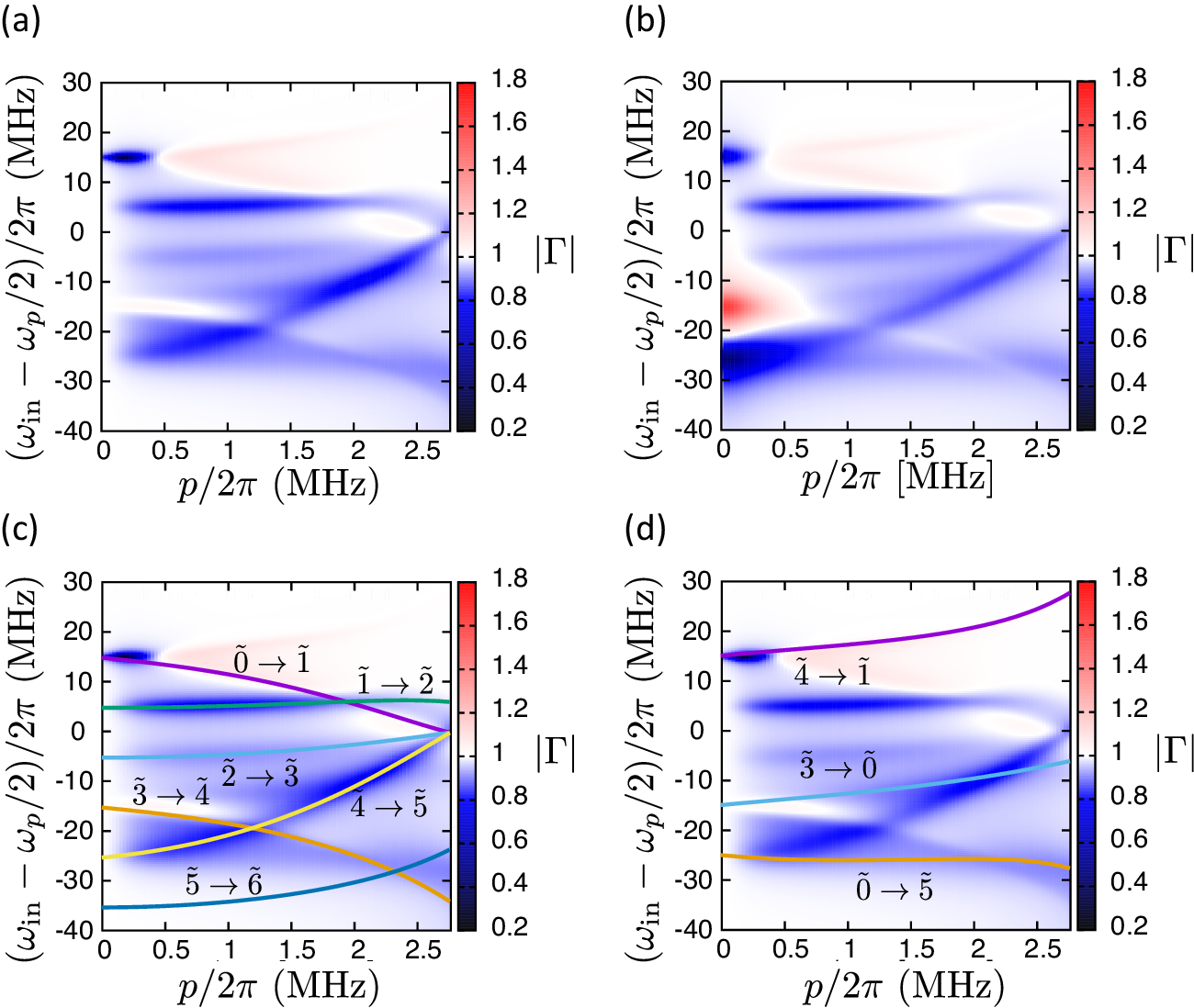}
\end{center}
\caption{
Amplitude of the reflection coefficient as a function of $\omega_{\rm in}$ and $p$ for a four-photon KPO in the stationary state.
The parameter set used is $K/2\pi=10$~MHz, $\Delta=1.5K$, $\kappa_{\rm ex}/2\pi=1$~MHz, and 
$\kappa_{\rm int}/2\pi=0.1~$MHz.
Panels (a) and (b)  are for the modified theory and the previous one, respectively.
Panels~(c) and (d) show the frequencies of relevant transitions overlaid on the data presented in panel (a).
}
\label{fourier_com_5_10_25}
\end{figure}

\begin{figure}[h!]
\begin{center}
\includegraphics[width=7.5cm]{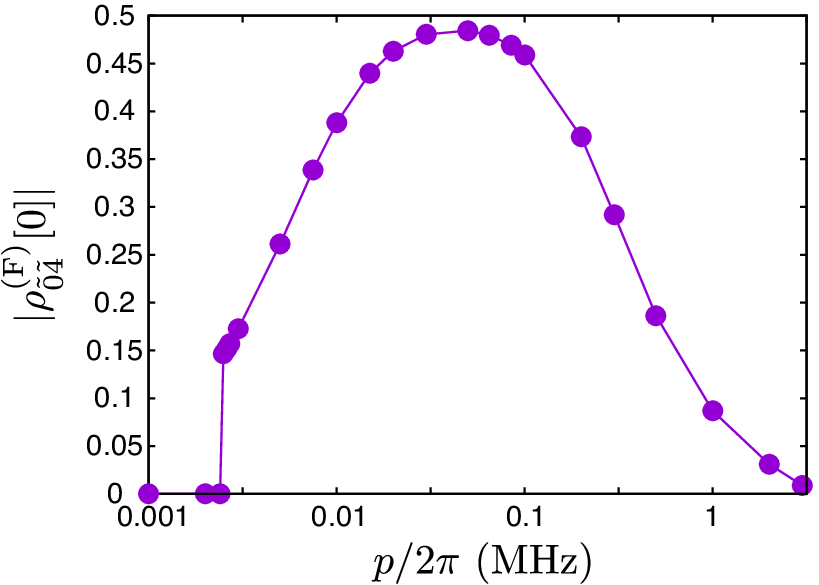}
\end{center}
\caption{
Amplitude of the density matrix element, $|\rho^{\rm (F)}_{\tilde{0}\tilde{4}}[0]|(=|\rho^{\rm (F)}_{\tilde{4}\tilde{0}}[0]|)$, of the stationary state as a function of $p$.
The parameters used are $K/2\pi=10$~MHz, $\Delta=1.5K$, $\kappa_{\rm ex}/2\pi=1$~MHz, and
$\kappa_{\rm int}/2\pi=0.1~$MHz. The solid line is shown as a guide to the eye.
}
\label{rho_data_5_16_25}
\end{figure}

The amplitude of the reflection coefficient $\Gamma$ is presented as a function of $\omega_{\rm in}$ and $p$ in figure~\ref{fourier_com_5_10_25}.
Figure~\ref{fourier_com_5_10_25}(a) and~\ref{fourier_com_5_10_25}(b) are for the modified theory and the previous one, respectively.  
In figure~\ref{fourier_com_5_10_25}(c,d), the transition frequencies between relevant levels are displayed with the same data as in figure~\ref{fourier_com_5_10_25}(a).
Clear differences in $|\Gamma|$ are observed, especially in the small-$p$ regime $(p/2\pi<1~$MHz).
In the previous theory displayed in figure~\ref{fourier_com_5_10_25}(b), there is a dip in $|\Gamma|$ erroneously deep at the frequency corresponding to the transition, $|\tilde{4}\rangle\rightarrow |\tilde{5}\rangle$, and there is a peak erroneously high at the frequency corresponding to the transition, $|\tilde{3}\rangle\rightarrow |\tilde{4}\rangle$, in contrast to the result of the modified theory.
These differences arise from the fact that the density matrix has considerably large off-diagonal elements in the small-$p$ regime. 
(The modified theory yields a result similar to that of the previous theory when we set $\rho^{\rm (F)}_{\tilde{l}\tilde{k}(\ne \tilde{l})}[0]=0$.)
Figure~\ref{rho_data_5_16_25} shows off-diagonal element, $|\rho^{\rm (F)}_{\tilde{0}\tilde{4}}[0]|$, as a function of $p$.
It is observed that $|\rho^{\rm (F)}_{\tilde{0}\tilde{4}}[0]|$ is large in the small-$p$ regime, except for very small values such as $p/2\pi<0.002~$MHz.
Although we observe an abrupt change in $|\rho^{\rm (F)}_{\tilde{0}\tilde{4}}[0]|$ at around $p/2\pi=0.002$~MHz, the origin of this behavior remains unclear.

\section{Transmission measurement}
We analyze the transmission measurement of a two-photon KPO, whose schematic is shown in figure~\ref{model_2_27_18}(b).
We derive the reflection and transmission coefficients based on the methods presented in Ref.~\cite{Masuda2021} and in section~\ref{Methods}.

The Hamiltonian of the system in a rotating frame at frequency $\omega_p/2$ is represented as
\begin{eqnarray}
\frac{\mathcal{H}}{\hbar} &=&  \frac{\mathcal{H}_{\rm KPO}^{(2)}}{\hbar}  +  \int_{0}^{\infty} dk \Big{[} (v_b k-\omega_p/2) b_{k}^{\dagger} b_{k} +
 \sqrt{\frac{v_b\kappa_{\rm ex}}{4\pi}}(a^\dagger b_{k} +  b_{k}^{\dagger} a) \Big{]}\nonumber\\
&& +  \int_{0}^{\infty} dk \Big{[} (v_b k-\omega_p/2) b_{k}'^{\dagger} b'_{k} +
 \sqrt{\frac{v_b\kappa_{\rm ex}}{4\pi}}(a^\dagger b'_{k} +  b_{k}'^{\dagger} a) \Big{]}\nonumber\\
&& +  \int_{0}^{\infty} dk \Big{[} (v_b k-\omega_p/2) c_{k}^{\dagger} c_{k} +
 \sqrt{\frac{v_b\kappa_{\rm int}}{4\pi}}(a^\dagger c_{k} +  c_{k}^{\dagger} a) \Big{]}\nonumber\\
&& +  \int_{0}^{\infty} dk \Big{[} (v_b k-\omega_p/2) c_{k}'^{\dagger} c'_{k} +
 \sqrt{\frac{v_b\kappa_{\rm int}}{4\pi}}(a^\dagger c'_{k} +  c_{k}'^{\dagger} a) \Big{]},
\label{H_T_4_21_23}
\end{eqnarray}
where $\mathcal{H}_{\rm KPO}^{(2)}$ is the Hamiltonian of the two-photon KPO given in equation~(\ref{H2KPO_6_24_25}).
Although we focus on a two-photon KPO as a specific example, the result can be applied to more general systems; for instance, $\mathcal{H}_{\rm KPO}^{(2)}$ can be replaced by the Hamiltonian of a multi-qubit system.
In equation~(\ref{H_T_4_21_23}), $b_k$ is the annihilation operator of a waveguide photon propagating forward with wave number $k>0$, while $b'_k$ corresponds to that propagating backward.
$c_k$ and $c'_k$ are the annihilation operators corresponding to internal loss.
The coupling terms include an additional factor of $1/\sqrt{2}$ compared to equation~(\ref{H_4_12_23}).
These factors are introduced to set the external and internal loss rates to $\kappa_{\rm ex}$ and $\kappa_{\rm int}$, respectively.

As shown in~\ref{Derivation of transmission and reflection coefficients},
the reflection and transmission coefficients are obtained as  
\begin{eqnarray}
\Gamma &=& - \frac{i}{E} \sqrt\frac{\kappa_{\rm ex}}{2v_b} \sum_{mn} X_{mn}\rho^{\rm (F)}_{nm}[-\tilde\omega_{\rm in}],\nonumber\\
T  &=& 1 - \frac{i}{E} \sqrt\frac{\kappa_{\rm ex}}{2v_b} \sum_{mn} X_{mn}\rho^{\rm (F)}_{nm}[-\tilde\omega_{\rm in}].
\label{T_Gamma_4_22_23}
\end{eqnarray}
These equations are the counterpart of equation~(\ref{Gamma_6_23_25}) for the reflection measurement.
The equation of motion for $\rho_{nm}$ is the same as equation~(\ref{drho_3_18_21}), although $\Omega$ is defined by 
\begin{eqnarray}
\Omega=\sqrt{\frac{v_b\kappa_{\rm ex}}{2}}E,
\label{Omega_4_22_23}
\end{eqnarray}
in this case.
The reflection and transmission coefficients satisfy $T=1+\Gamma$.

\subsection{Analytic formulas without interference}
We derive analytic formulas for the reflection and transmission coefficients in the case where the interference effect and the off-diagonal elements $\rho^{\rm (F)}_{lk(\ne l)}[0]$ are negligible. These contributions are incorporated in the following subsection.

We assume that the probe field is nearly resonant only with the transition $|\phi_m\rangle \rightarrow |\phi_n\rangle$, that is, $\omega_{\rm in} - \omega_p/2 + \omega_m - \omega_n \simeq 0$.
The off-diagonal element of the density matrix $\rho^{\rm (F)}_{lk}[-\tilde\omega_{\rm in}]$ is nonzero only when $(k,l)=(m,n)$, while non-resonant elements such as $\rho^{\rm (F)}_{lk}[-2\tilde\omega_{\rm in}]$ are zero.
We further assume that the presence of the weak probe field does not affect zero-frequency diagonal elements, $\rho^{\rm (F)}_{mm}[0]$, and thus $\rho^{\rm (F)}_{mm}[0]$ is taken to be the same as in the stationary state without the probe field.
Then,  equation~(\ref{rho_F_v1_2_20_21}) yields $\rho^{\rm (F)}_{nm}[-\tilde\omega_{\rm in}]$ expressed in equation~(\ref{rhomn_2_19_21}).
Using equation~(\ref{rhomn_2_19_21}) in equation~(\ref{T_Gamma_4_22_23}), we obtain the following expressions:
\begin{eqnarray}
\Gamma &=& \frac{1}{2} \sum_{mn} \xi_{mn},\nonumber\\
T &=& 1+ \frac{1}{2} \sum_{mn} \xi_{mn},
\end{eqnarray}
with 
\begin{eqnarray}
\xi_{mn}  =  \frac{\kappa_{\rm ex} |X_{mn}|^2 (\rho^{\rm (F)}_{mm}[0] - \rho^{\rm (F)}_{nn}[0]) }{i\Delta_{nm}- \frac{\kappa_{\rm ex}+\kappa_{\rm int}}{2}(Y_{mm}+Y_{nn})}.
\label{Gamma_4_22_23}
\end{eqnarray}

The nominal loss rates corresponding to the spectrum of the transition, $|\phi_m\rangle \rightarrow |\phi_n\rangle$, are defined by comparing the reflection coefficient, $\Gamma=\xi_{mn}/2$, with that of a linear resonator represented as $\Gamma_r=(\kappa_{\rm ex}^{(r)}/2)/[i\Delta_r - (\kappa_{\rm ex}^{(r)} + \kappa_{\rm int}^{(r)})/2]$.
The nominal external and internal loss rates are written as 
\begin{eqnarray}
\tilde\kappa^{(mn)}_{\rm ex} &=& \kappa_{\rm ex} |X_{mn}|^2 (\rho^{\rm (F)}_{mm}[0] - \rho^{\rm (F)}_{nn}[0]),\nonumber\\
\tilde\kappa^{(mn)}_{\rm int} &=& (\kappa_{\rm ex}+\kappa_{\rm int})(Y_{mm}+Y_{nn}) - \kappa_{\rm ex} |X_{mn}|^2 (\rho^{\rm (F)}_{mm}[0] - \rho^{\rm (F)}_{nn}[0]).
\label{nominal_4_23_23}
\end{eqnarray}
These are identical to those for reflection measurement in reference~\cite{Masuda2021}. 

\subsection{Analytic formulas with interference}
When the interference effect and zero-frequency off-diagonal elements, $\rho^{\rm (F)}_{lk(\ne l)}[0]$, are not negligible, 
$\rho^{\rm (F)}_{nm}[-\tilde\omega_{\rm in}]/\Omega$ can be obtained by solving the simultaneous equation~(\ref{rho_nm_2_16_22}).
The reflection and transmission coefficients are obtained by substituting it into equation~(\ref{T_Gamma_4_22_23}).

We focus on the case where the two ground states, $|\tilde{0}\rangle$ and $|\tilde{1}\rangle$, and a pair of excited states, $|\tilde{2}\rangle$ and $|\tilde{3}\rangle$, are each degenerate, as we did in section~\ref{Analytic results}.
Substitution of equation~(\ref{rho_two_7_31_25}) into equation~(\ref{T_Gamma_4_22_23}) leads to
\begin{eqnarray}
\Gamma &=& -i\frac{\kappa_{\rm ex}}{2A} \Big{[}
X_{\tilde{0}\tilde{3}} (a_{22}b_1-a_{12}b_2)
+ X_{\tilde{1}\tilde{2}}( a_{11}b_2-a_{21}b_1)  \Big{]},\nonumber\\
T &=& 1-i\frac{\kappa_{\rm ex}}{2A} \Big{[}
X_{\tilde{0}\tilde{3}} (a_{22}b_1-a_{12}b_2)
+ X_{\tilde{1}\tilde{2}}( a_{11}b_2-a_{21}b_1)  \Big{]},
\label{Gamma_t_0312_8_6_25}
\end{eqnarray}
where $a_{ij}$ and $b_i$ are defined just below equation~(\ref{eq_a_b_7_31_25}), and $A=a_{11}a_{22}-a_{12}a_{21}$.

For sufficiently large pump amplitude, we can further simplify $\Gamma$ and $T$ as
\begin{eqnarray}
\Gamma &=& \frac{\kappa_{\rm ex}}{2(i\tilde{\Delta}-\frac{\kappa_{\rm ex}+\kappa_{\rm int}}{2})},\nonumber\\
T &=& 1 + \frac{\kappa_{\rm ex}}{2(i\tilde{\Delta}-\frac{\kappa_{\rm ex}+\kappa_{\rm int}}{2})},
\label{Gamma_ana_t_7_24_25}
\end{eqnarray}
where we used the explicit expressions for the energy eigenstates, given just below equation~(\ref{Gamma_0312_7_24_25}).
There is no $\alpha$ dependence in $\Gamma$ and $T$ in the large-pump-amplitude limit, and they have the same form as those for a linear resonator.

\section{Conclusion}
We have developed a theory for reflection and transmission measurements of driven nonlinear resonators, focusing on two-photon and four-photon KPOs.
The previous theory has been modified to incorporate the interference between inter-level transitions and the zero-frequency off-diagonal elements of the density matrix.
The spectrum exhibits clear signatures of inter-transition interference when the transitions are energetically close.
We have clarified the conditions under which the interference effect arises and the conditions under which the off-diagonal elements affect the spectrum.
Furthermore, we have analytically and numerically analyzed the interference effect, and derived analytic formulas for the reflection and transmission coefficients under typical two-photon KPO parameters.

\section*{Acknowledgements}
This paper is based on results obtained from a project, JPNP16007, commissioned by the New Energy and Industrial Technology Development Organization (NEDO), Japan.


\section*{Data availability statement}
All data that support the findings of this study are included within the article.

\appendix

\setcounter{equation}{0}
\renewcommand*{\theequation}{A\arabic{equation}}
\renewcommand*{\thesection}{Appendix \Alph{section}}

\section{Derivation of transmission and reflection coefficients}
\label{Derivation of transmission and reflection coefficients}
We derive the transmission and reflection coefficients for the transmission measurement in the manner used for the reflection measurement in Ref.~\cite{Masuda2021} (see also ~\cite{Koshino2012, Yamamoto2016, Banacloche2013}).
In the lab frame, the Hamiltonian of the system is represented as
\begin{eqnarray}
\frac{\mathcal{H}}{\hbar} &=&  \frac{\mathcal{H}_{\rm KPO}^{(2,L)}}{\hbar}  +  \int_{0}^{\infty} dk \Big{[} v_b k b_{k}^{\dagger} b_{k} +
 \sqrt{\frac{v_b\kappa_{\rm ex}}{4\pi}}(a^\dagger b_{k} +  b_{k}^{\dagger} a) \Big{]}\nonumber\\
&& +  \int_{0}^{\infty} dk \Big{[} v_b k b_{k}'^{\dagger} b'_{k} +
 \sqrt{\frac{v_b\kappa_{\rm ex}}{4\pi}}(a^\dagger b'_{k} +  b_{k}'^{\dagger} a) \Big{]}\nonumber\\
&& +  \int_{0}^{\infty} dk \Big{[} v_b k c_{k}^{\dagger} c_{k} +
 \sqrt{\frac{v_b\kappa_{\rm int}}{4\pi}}(a^\dagger c_{k} +  c_{k}^{\dagger} a) \Big{]}\nonumber\\
&& +  \int_{0}^{\infty} dk \Big{[} v_b k c_{k}'^{\dagger} c'_{k} +
 \sqrt{\frac{v_b\kappa_{\rm int}}{4\pi}}(a^\dagger c'_{k} +  c_{k}'^{\dagger} a) \Big{]},
\label{H_t_L_9_8_23}
\end{eqnarray}
where $\mathcal{H}_{\rm KPO}^{(2,L)}$ is the Hamiltonian of the two-photon KPO in the lab frame.
We refer readers interested in an explicit form of $\mathcal{H}_{\rm KPO}^{(2,L)}$ to Ref.~\cite{Masuda2021}.
The Heisenberg equation of motion of $b_{k}$ corresponding to Hamiltonian (\ref{H_T_4_21_23}) is represented as
\begin{eqnarray}
\frac{d}{dt}b_{k}(t) = -i  v_b k  b_{k}(t) - i\sqrt{\frac{v_b \kappa_{\rm ex}}{4\pi}} a(t).
\label{Hei_4_21_23}
\end{eqnarray}
A formal solution of equation~(\ref{Hei_4_21_23}) is written as
\begin{eqnarray}
b_{k}(t) = b_{k}(0)e^{-i kv_b t} - i\sqrt{\frac{v_b \kappa_{\rm ex}}{4\pi}}\int_0^t d\tau a(\tau) e^{i kv_b(\tau-t)},
\label{blk_4_21_23}
\end{eqnarray}
for $t>0$.
We extend the lower limit of $k$ to $-\infty$ in order to introduce the real-space representation
of the field operator defined by
\begin{eqnarray}
\tilde{b}_{r}(t) = \frac{1}{\sqrt{2\pi}} \int_{-\infty}^{\infty}dk e^{ikr}b_{k}(t),
\label{br_4_21_23}
\end{eqnarray}
for $-\infty < r < \infty$.

The negative and positive regions are assigned to the incoming and outgoing fields, respectively, where $r$ is regarded as a virtual coordinate.
The input field operator $\tilde{b}_{r}^{(\rm in)}$ and the output field operator $\tilde{b}_{r}^{(\rm out)}$ are defined by 
\begin{eqnarray}
\tilde{b}_{r}^{(\rm in)}(t) = \tilde{b}_{-r}(t),\nonumber\\
\tilde{b}_{r}^{(\rm out)}(t) = \tilde{b}_{r}(t),
\label{b_in_out_4_21_23}
\end{eqnarray}
for $r\ge 0$.
Equations~(\ref{blk_4_21_23}) and (\ref{br_4_21_23}) lead to  
\begin{eqnarray}
\tilde{b}_{r}(t) = \tilde{b}_{r-v_bt}(0) - i \sqrt{\frac{\kappa_{\rm ex}}{2v_b}} \theta(r)\theta(v_bt-r) a(t-r/v_b),
\label{br_4_21_23_2}
\end{eqnarray}
where $ \theta$ is the Heaviside step function.
Also, we used 
\begin{eqnarray}
\frac{1}{2\pi}\int_{-\infty}^\infty dk e^{ik[r+v_b(\tau-t)]} = \delta(r+v_b(\tau-t)) = \frac{1}{v_b} \delta(r/v_b+\tau-t).
\end{eqnarray}
The step functions originate from an integration of the delta function.
Equation~(\ref{b_in_out_4_21_23}) for $r=0$ and equation~(\ref{br_4_21_23}) lead to the relation,
\begin{eqnarray}
\tilde{b}_{0}(t) = \tilde{b}_{v_bt}^{(\rm in)}(0) - \frac{i}{2}\sqrt{\frac{\kappa_{\rm ex}}{2v_b}} a(t),
\label{in_out_4_21_23}
\end{eqnarray}
where we used $\theta(0)=1/2$.
We have similar relations for $\tilde{b}'_{0}(t)$, $\tilde{c}_{0}(t)$, $\tilde{c}'_{0}(t)$.

We assume that an input microwave is applied from one side of the TL.
The density operator of the initial state is assumed to be given by
\begin{eqnarray}
\rho_0 = \rho^{\rm (L)} \otimes |\Psi_{\rm tl}\rangle \langle \Psi_{\rm tl} |,
\end{eqnarray}
with
\begin{eqnarray}
|\Psi_{\rm tl}\rangle = \mathcal{N}\exp\Big{[}\int_{-\infty}^{0}dr E_{\rm in}(-r) \tilde{b}_{r}^\dagger(0) \Big{]} |v_{\rm tl}\rangle.
\label{Psi_tl_4_21_23}
\end{eqnarray}
where $|v_{\rm tl}\rangle$ is the vacuum state of the TL modes, and $\rho^{\rm (L)}$ is the density operator of the KPO in the lab frame.
The TL is in a continuous-mode coherent state.
Here, $E_{\rm in}$ is given by 
\begin{eqnarray}
E_{\rm in}(r) = \left\{
\begin{array}{cl}
E e^{- i \omega_{\rm in} r/v_b} & (r>0) \\
0 & ({\rm otherwise}),
\end{array}
\right.
\label{E_9_29_20}
\end{eqnarray}
where $E$ and $\omega_{\rm in}$ are the amplitude and the angular frequency of the incoming microwave, respectively.
We obtain
\begin{eqnarray}
\tilde{b}_{v_bt}^{({\rm in})}(0) |\Psi_{\rm tl}\rangle = E_{\rm in}(v_bt) |\Psi_{\rm tl}\rangle 
= E e^{-i\omega_{\rm in}t}  |\Psi_{\rm tl}\rangle
\label{blvt_4_21_23}
\end{eqnarray}
as a consequence of equations~(\ref{Psi_tl_4_21_23}) and (\ref{E_9_29_20}).

We substitute $r=+0$ in equation~(\ref{br_4_21_23_2}) and take the expectation value with respect to $\rho_0$ to obtain the input-output relation
\begin{eqnarray}
\langle \tilde{b}_{+0}^{(\rm out)}(t) \rangle = \langle \tilde{b}_{v_bt}^{(\rm in)}(0) \rangle 
- i\sqrt\frac{\kappa_{\rm ex}}{2v_b} \langle a(t) \rangle.
\label{in_out_4_21_23_2}
\end{eqnarray}
Equation~(\ref{blvt_4_21_23}) leads to 
\begin{eqnarray}
\langle \tilde{b}_{v_bt}^{(\rm in)}(0) \rangle = E e^{-i\omega_{\rm in}t}.
\label{Bin_4_21_23}
\end{eqnarray}

Our analysis focuses on the Fourier component of $\langle \tilde{b}_{+0}^{(\rm out)}(t) \rangle$ with a frequency of $-\omega_{\rm in}$, identical to that of the input field.
We define the transmission coefficient as 
\begin{eqnarray}
T = \langle \tilde{b}_{+0}^{(\rm out)} \rangle[-\omega_{\rm in}]  / E,
\label{T_out_4_21_23}
\end{eqnarray}
where 
$\langle \tilde{b}_{+0}^{(\rm out)} \rangle[-\omega_{\rm in}]$ is the Fourier component of $\langle \tilde{b}_{+0}^{(\rm out)} (t) \rangle$ with a frequency of $-\omega_{\rm in}$.
Equation (\ref{T_out_4_21_23}) can be rewritten using equations~(\ref{in_out_4_21_23_2}) and (\ref{Bin_4_21_23}) as
\begin{eqnarray}
T = 1 - \frac{i}{E}\sqrt{\frac{\kappa_{\rm ex}}{2v_b}} \langle a\rangle [-\omega_{\rm in}],
\label{Trans_4_21_23}
\end{eqnarray}
where $ \langle a\rangle [-\omega_{\rm in}]$ is the Fourier component of $\langle a(t)\rangle$ with a frequency of $-\omega_{\rm in}$.

Similarly we have the relation for mode propagating backward: 
\begin{eqnarray}
\langle \tilde{b}_{+0}'^{(\rm out)}(t) \rangle = \langle \tilde{b}_{v_bt}'^{(\rm in)}(0) \rangle 
- i\sqrt\frac{\kappa_{\rm ex}}{2v_b} \langle a(t) \rangle.
\label{back_in_out_4_21_23}
\end{eqnarray}
Here, the first term is zero due to equation~(\ref{Psi_tl_4_21_23}).
The reflection coefficient is defined by
\begin{eqnarray}
\Gamma &=& \langle \tilde{b}_{+0}'^{(\rm out)} \rangle[-\omega_{\rm in}]  / E\nonumber\\
&=& - \frac{i}{E}\sqrt\frac{\kappa_{\rm ex}}{2v_b} \langle a \rangle  [-\omega_{\rm in}].
\label{Gamma_dash_4_21_23}
\end{eqnarray}

We can rewrite $ \langle a(t)\rangle$ as
\begin{eqnarray}
\langle a(t)\rangle={\rm Tr}[a\rho^{(L)}(t)]  &=& \sum_{m} \langle \phi_m | a\rho^{(L)}(t) |\phi_m \rangle = \sum_{mn} X_{mn}\rho^{(L)}_{nm}(t),
\label{A_4_22_23}
\end{eqnarray}
where $\rho^{(L)}_{nm} = \langle \phi_n | \rho^{(L)} | \phi_m \rangle$, and 
$X_{mn} = \langle \phi_m | a | \phi_n \rangle$.
Therefore, we have $\langle a\rangle [-\omega_{\rm in}]= \sum_{mn} X_{mn}\rho^{\rm (F)}_{nm} [-\omega_{\rm in}]$.
We transform to a rotating frame at frequency $\omega_p/2$.
In this frame, the reflection and transmission coefficients are determined by $\langle a\rangle [-\tilde\omega_{\rm in}]$ rather than by $\langle a\rangle [-\omega_{\rm in}]$, where $\tilde\omega_{\rm in}=\omega_{\rm in}-\omega_p/2$. 
We have 
\begin{eqnarray}
\langle a\rangle [-\tilde\omega_{\rm in}]= \sum_{mn} X_{mn}\rho^{\rm (F)}_{nm} [-\tilde\omega_{\rm in}],
\label{a_omega_7_14_25}
\end{eqnarray}
By using equation~(\ref{Trans_4_21_23}),~(\ref{Gamma_dash_4_21_23}),~(\ref{a_omega_7_14_25}), we obtain the transmission and reflection coefficients as equation~(\ref{T_Gamma_4_22_23}).

In the rotating frame, the equation of motion for the KPO is written as
\begin{eqnarray}
\frac{d\rho}{dt} = -\frac{i}{\hbar} [\tilde{\mathcal{H}}^{(2)}_{\rm KPO},\rho] +  \mathcal{L}[\rho],
\label{ME_4_21_23}
\end{eqnarray}
with 
\begin{eqnarray}
\frac{\tilde{\mathcal{H}}^{(2)}_{\rm KPO}}{\hbar} &=& \frac{\mathcal{H}^{(2)}_{\rm KPO}}{\hbar} + \sqrt{\frac{v_b \kappa_{\rm ex}}{2}}(E e^{i \tilde\omega_{\rm in} t} a + Ee^{-i\tilde\omega_{\rm in} t} a^\dagger),\nonumber\\
\mathcal{L}[\rho] &=& -\frac{\kappa_{\rm tot}}{2} (a^\dagger a \rho + \rho a^\dagger a - 2a\rho a^\dagger),
\label{L_7_14_25}
\end{eqnarray}
where $\tilde{\mathcal{H}}^{(2)}_{\rm KPO}$ includes the effect of the probe field, and $\mathcal{H}^{(2)}_{\rm KPO}$ is given in equation~(\ref{H2KPO_6_24_25}).
By using equations (\ref{ME_4_21_23}) and (\ref{L_7_14_25}), we can obtain the equation of motion (\ref{rho_F_v1_2_20_21}) for the Fourier component of the density matrix elements.

\end{document}